\date{October 31, 2012}

\documentclass[a4paper]{jpconf}

\usepackage{epsfig}

\usepackage{float}

\begin{document}

\title{Galactic rotation curves in conformal gravity}

\author{Philip D Mannheim$^1$ and James G O'Brien$^2$}
\address{$^1$Department of Physics, University of Connecticut, Storrs, CT 06269, USA}
\address{$^2$Department of Sciences, Wentworth Institute of Technology, Boston, MA 02115, USA}
\ead{$^1$philip.mannheim@uconn.edu}
\ead{$^2$obrienj10@wit.edu}

\begin{abstract}
We review some recent work by Mannheim and O'Brien on the systematics of galactic  rotation curves in the conformal gravity theory.  In this work  the conformal theory was applied to a comprehensive, high quality sample of spiral galaxies whose rotation curves extend well beyond the galactic optical disks.  On galactic scales the conformal gravitational theory departs from the standard Newtonian theory in two distinct ways. One is a local way in which local matter sources within galaxies generate not just  Newtonian potentials but  linear potentials as well.  The other is a global way in which  two universal global potentials, one linear the other quadratic, are generated by the rest of the matter in the universe.  The  study involves a broad set of 138 spiral galaxies of differing luminosities and sizes, and is augmented here through the inclusion of an additional three tidal dwarf galaxies.  With its linear and quadratic potentials the conformal theory can account for the systematics of an entire 141 galaxy sample without any need for galactic dark matter, doing so with only one free parameter per galaxy, namely the visible galactic mass to light ratio.   
\end{abstract}

\section{Introduction}
\label{s1}
\medskip

Over the past three decades, the persistence of the missing mass or dark matter problem has generated an increasing interest in alternative gravitational theories.  In the time since the classic studies (see e.g. \cite{Rubin2000}) of the missing mass problem in spiral galaxies galactic observational techniques have improved, and it has become possible to study and constrain the motions of luminous matter in galaxies with great precision.  With rotational data for spiral galaxies now extending well beyond galactic optical disks, one finds that in essentially every case studied the measured rotational velocities do not conform with the familiar Newtonian gravity expectation associated with the observed visible material.  It is important to note that the Newtonian  expectation on galactic distance scales is derived by extrapolating standard gravity beyond solar system distance scales without any modification.  To determine  the expected Newtonian prediction  one treats the galaxy as a collection of $N^*$ individual sources each of  typical  mass $M_\odot$ and combines the Newtonian gravitational potentials generated by each of the individual sources.  For a thin disk-shaped galaxy  with a typical exponential surface brightness distribution $\Sigma(R)=\Sigma_0e^{-R/R_0}$ with scale length $R_0$, the resulting circular velocity for a test particle at a radial distance $R$ from the center of the disk  is given by the Freeman formula  (see e.g. \cite{Mannheim2006})
\begin{equation}
v^2(R)=
\frac{N^*M_\odot G  R^2}{2R_0^3}\left[I_0\left(\frac{R}{2R_0}
\right)K_0\left(\frac{R}{2R_0}\right)-
I_1\left(\frac{R}{2R_0}\right)
K_1\left(\frac{R}{2R_0}\right)\right].
\label{free}
\end{equation}
The failure of equation (\ref{free}) using known visible matter sources alone can be alleviated via the ad hoc addition of dark, non-luminous, mass sources, as they can then bring the theory into agreement  with data, and use of such dark matter sources has  been extensively studied in the literature. However, the ability of such sources to fit data requires that the distribution of these sources be specifically chosen in each given galaxy on a case by case  basis rather than in a way that is universal, with a substantial amount of dark matter always needing to be located in the outer regions of galaxies where there is little or no visible matter.  Since so many galaxies over so extensive a range of sizes and luminous distributions have now been found  to have a missing mass problem,  some authors have considered the inability of luminous sources to fit the data on their own to be indicative of a failure of the Newtonian theory on galactic distance scales rather than of the presence of dark matter.  Hence, these authors have suggested that is the extrapolation of solar system Newtonian gravity to galactic distance scales that is the cause of the problem (see e.g. \cite{Mannheim2006} for a recent review).  Various candidate alternative theories have been presented in the literature such as  Milgrom's Modified Newtonian Dynamics (MOND)  \cite{Milgrom1983a} and Moffat's Metric Skew Tensor Gravity (MSTG) \cite{Moffat2005}.  It is the purpose of this review to highlight recent work done by Mannheim and O'Brien in applying another candidate alternative theory, namely the conformal gravity theory, to a large data sample containing some of the most well-studied spiral galaxies in the astronomical literature.  Using the best galactic optical and radio data available, and a standardized, non-biased, treatment for selecting input galactic parameters, we find that the conformal theory is able to provide for a good accounting of the data without the need for any dark matter whatsoever.  As with both MOND and MSTG, the conformal theory provides a universal prescription for departures from standard Newtonian gravity, to thus require no galaxy by galaxy dependent free parameters other than the standard galactic visible  mass to light ratio common to all theories of galactic rotation curves.

The conformal theory of gravity was originally formulated by Weyl shortly after the advent of General Relativity. As such, it is a  completely covariant metric theory of gravity, as endowed with an additional symmetry, namely local conformal invariance.  This conformal symmetry forbids the presence of both any fundamental Einstein-Hilbert action term (with its dimensionful Newton constant) or any fundamental cosmological constant term in the gravitational action, with the gravitational action being uniquely prescribed to be of the form 
\begin{equation}
I_{\rm W}=-\alpha_g\int d^4x\, (-g)^{1/2}C_{\lambda\mu\nu\kappa}
C^{\lambda\mu\nu\kappa}
\equiv -2\alpha_g\int d^4x\, (-g)^{1/2}\left[R_{\mu\kappa}R^{\mu\kappa}-\frac{1}{3} (R^{\alpha}_{\phantom{\alpha}\alpha})^2\right].
\label{182}
\end{equation}
In equation (\ref{182}) the the gravitational coupling constant $\alpha_g$ is dimensionless and the tensor $C_{\lambda\mu\nu\kappa}$ defined by 
\begin{equation}
C_{\lambda\mu\nu\kappa}= R_{\lambda\mu\nu\kappa}
-\frac{1}{2}\left(g_{\lambda\nu}R_{\mu\kappa}-
g_{\lambda\kappa}R_{\mu\nu}-
g_{\mu\nu}R_{\lambda\kappa}+
g_{\mu\kappa}R_{\lambda\nu}\right)
+\frac{1}{6}R^{\alpha}_{\phantom{\alpha}\alpha}\left(
g_{\lambda\nu}g_{\mu\kappa}-
g_{\lambda\kappa}g_{\mu\nu}\right)
\label{180}
\end{equation}
is the conformal Weyl tensor.  The conformal action is invariant under local conformal transformations of the metric of the form $g_{\mu\nu}(x)\rightarrow e^{2\alpha(x)}g_{\mu\nu}(x)$ where $\alpha(x)$ is a local, spacetime-dependent phase.  Because of its  scale invariance, the conformal theory does not suffer from the cosmological problem present in the standard Einstein gravitational theory \cite{Mannheim2011a,Mannheim2011b}.

Unlike the standard Einstein theory where the action is linear in the Riemann curvature tensor, the conformal action is  quadratic in the Riemann tensor. Consequently, its equations of motion are fourth-order derivative functions of the metric, and in the presence of a matter source $T^{\mu\nu}$ take the form  
\begin{equation}
4\alpha_g W^{\mu\nu}=4\alpha_g\left[
2C^{\mu\lambda\nu\kappa}_
{\phantom{\mu\lambda\nu\kappa};\lambda;\kappa}-
C^{\mu\lambda\nu\kappa}R_{\lambda\kappa}\right]=4\alpha_g\left[W^{\mu
\nu}_{(2)}-\frac{1}{3}W^{\mu\nu}_{(1)}\right]=T^{\mu\nu},
\label{188}
\end{equation}
where the tensors $W^{\mu\nu}_{(1)}$ and $W^{\mu\nu}_{(2)}$ are given by 
\begin{eqnarray}
&&W^{\mu \nu}_{(1)}=
2g^{\mu\nu}(R^{\alpha}_{\phantom{\alpha}\alpha})          
^{;\beta}_{\phantom{;\beta};\beta}                                              
-2(R^{\alpha}_{\phantom{\alpha}\alpha})^{;\mu;\nu}                           
-2 R^{\alpha}_{\phantom{\alpha}\alpha}
R^{\mu\nu}                              
+\frac{1}{2}g^{\mu\nu}(R^{\alpha}_{\phantom{\alpha}\alpha})^2,
\nonumber \\
&&W^{\mu \nu}_{(2)}=
\frac{1}{2}g^{\mu\nu}(R^{\alpha}_{\phantom{\alpha}\alpha})   
^{;\beta}_{\phantom{;\beta};\beta}+
R^{\mu\nu;\beta}_{\phantom{\mu\nu;\beta};\beta}                     
 -R^{\mu\beta;\nu}_{\phantom{\mu\beta;\nu};\beta}                        
-R^{\nu \beta;\mu}_{\phantom{\nu \beta;\mu};\beta}                          
 - 2R^{\mu\beta}R^{\nu}_{\phantom{\nu}\beta}                                    
+\frac{1}{2}g^{\mu\nu}R_{\alpha\beta}R^{\alpha\beta}.
\label{EX1}
\end{eqnarray}

Despite its somewhat formidable appearance, Mannheim and Kazanas \cite{Mannheim1989,Mannheim1994} were able to find a closed form analytic solution to equation (\ref{188}) in the case of a standard static spherically symmetric geometry with line element 
\begin{equation}
ds^2=B(r)c^2dt^2-A(r)dr^2-r^2d\Omega_2,
\label{EX2}
\end{equation}
where $d\Omega_2=d\theta^2+\sin^2\theta d\phi^2$. Specifically, using the conformal symmetry they showed that one could transform this line element into one that was conformal to a line element with $A(r)=1/B(r)$, with equation (\ref{188}) then reducing (up to an irrelevant conformal transformation) to the remarkably simple
\begin{equation}
\nabla^4B(r)=B^{\prime\prime\prime\prime}+\frac{4B^{\prime\prime\prime}}{r}=\frac{3}{4\alpha_gB(r)}(T^{0}_{\phantom{0}0}-T^{r}_{\phantom{r}r})\equiv f(r),
\label{EX3}
\end{equation}
a relation that is exact without approximation. The general solution to equation (\ref{EX3})  is given by 
\begin{eqnarray}
B(r)&=& -\frac{r}{2}\int_0^r
dr^{\prime}\,r^{\prime 2}f(r^{\prime})
-\frac{1}{6r}\int_0^r
dr^{\prime}\,r^{\prime 4}f(r^{\prime})
\nonumber\\
&&
-\frac{1}{2}\int_r^{\infty}
dr^{\prime}\,r^{\prime 3}f(r^{\prime})
-\frac{r^2}{6}\int_r^{\infty}
dr^{\prime}\,r^{\prime }f(r^{\prime})+\hat{B}(r),
\label{EX4}
\end{eqnarray}                                 
where $\hat{B}(r)$ obeys $\nabla^4\hat{B}(r)=0$. Solutions to equation (\ref{EX3}) will depend on the matter distribution $f(r)$, and unlike the standard second-order theory where only local matter inside of a source generates a gravitational force, in the conformal theory both local matter inside a source (the first two integrals  in equation (\ref{EX4})) and the global matter exterior to it (the third and fourth integrals in equation (\ref{EX4})) can make contributions  \cite{Mannheim1997}. Thus to determine  motions within galaxies one needs to consider local contributions coming from the visible material inside the galaxy and global contributions coming from material outside the galaxy (viz. the rest of the universe). As we will see below, the combined effect of these local and global contributions  will enable us to provide an accounting of galactic rotation curve data without the need to invoke dark matter.

Given equation (\ref{EX4}), a star of radius $r_0$  and source function $f^*(r)$ will generate a gravitational potential of the form
\begin{equation}
V^{*}(r>r_0)=-\frac{\beta^*c^2}{r}+\frac{\gamma^{*} c^2 r}{2}
\label{E4}
\end{equation}
(as normalized to unit solar mass units), where
\begin{equation}
\gamma^*= -\frac{1}{2}\int_0^{r_0}dr^{\prime}\,r^{\prime 2}f^*(r^{\prime}),\qquad 
2\beta^*=\frac{1}{6}\int_0^{r_0}\,dr^{\prime}\,r^{\prime 4}f^*(r^{\prime}).
\label{E3}
\end{equation}
A star such as the sun thus puts out both a Newtonian potential $-\beta^*c^2/r$ and a linear potential $\gamma^*c^2r/2$. As can be seen, on identifying $\beta^* =M_{\odot}G/c^2$ the potential given in equation (\ref{E4}) reduces to the standard  Newtonian potential  in the small $r$ limit where the $\gamma^* r$ term becomes negligible.  However, for large enough $r$ this same term would become competitive with the Newtonian  term, to thereby precisely provide for our desired departure from standard Newtonian gravity on large distance scales.

For applications to galaxies, we need to integrate $V^*(r)$ over the local matter distribution in the galaxy. On taking the galaxy to be a thin disk with $N^*$ stars distributed with surface brightness  $\Sigma (R)=\Sigma_0e^{-R/R_0}$ and luminosity $L=2\pi R_0^2 \Sigma_0$,  the net contribution to circular velocities due the local material within galaxies is found to be given by \cite{Mannheim2006}
\begin{eqnarray}
v^2_{\rm LOC}({\rm disk})&=&\frac{N^*\beta^*c^2 R^2}{2R_0^3}\left[I_0\left(\frac{R}{2R_0}
\right)K_0\left(\frac{R}{2R_0}\right)-
I_1\left(\frac{R}{2R_0}\right)
K_1\left(\frac{R}{2R_0}\right)\right] 
\nonumber \\
&+& \frac{N^*\gamma^* c^2R^2}{2R_0}I_1\left(\frac{R}{2R_0}\right)
K_1\left(\frac{R}{2R_0}\right).
\label{E5}
\end{eqnarray}
The ratio $N^*M_{\odot}/L$ is recognized as the galactic mass to light ratio $M/L$.
Equation (\ref{E5}) generalizes the Newtonian expectation given in equation (\ref{free}) and gives the net conformal gravity contribution to circular velocities due to the local material within a galaxy.

As regards the global effect due to the rest of the material in the universe, we note that cosmologically there are two main global components, a cosmological background that is homogeneous and isotropic, and  inhomogeneities in it that arise due to fluctuations around that background. In the conformal theory both of these components make a contribution to motions within galaxies, but they do so in quite different ways. Since the cosmological background is both homogeneous and isotropic, it can be described by a Robertson-Walker (RW) geometry with isotropic coordinate line element 
\begin{equation}
ds^2=c^2d\tau^2-\frac{R^2(\tau)}{[1+K\rho^2/4]^2}
\left(d\rho^2+\rho^2d\Omega_2\right).
\label{EX5}
\end{equation}                                 
With an RW geometry being conformal to flat, in it  both the Weyl tensor  $C^{\lambda\mu\nu\kappa}$ and the $W^{\mu\nu}$ tensor introduced in equation (\ref{188}) vanish identically. In consequence the background matter energy-momentum tensor must vanish also (something it both can do and does do non-trivially  in the conformal case \cite{Mannheim2006}), with the cosmic background thus only contributing to the $\hat{B}(r)$ term in equation (\ref{EX4}). Since the presence of inhomogeneities  breaks the conformal to flat structure of the background geometry, inhomogeneities then lead to both a non-vanishing $W^{\mu\nu}$ and a non-vanishing $f(r)$, and hence contribute to the global third and fourth integrals in equation (\ref{EX4}).  Moreover, no matter in what specific way the cosmological background and the inhomogeneities in it might contribute to equation (\ref{EX4}), they must do so in a way that is independent of any given galaxy, i.e. they must contribute universally via an effect that is to act in exactly the same way on each and every galaxy, without any regard to their morphologies, sizes, luminosities, or masses. We thus anticipate, and will in fact find, that there must be some universality in the galactic rotation curve data.

To determine the specific contribution of  the cosmological background to galactic motions we need to transform the comoving cosmological RW geometry to the Schwarzschild coordinate rest frame coordinate system given in (\ref{EX2}). To this end we note \cite{Mannheim1989} that the coordinate transformation  
\begin{equation}
 \rho=\frac{4r}{[2(1+\gamma_0r)^{1/2}+2 +\gamma_0 r]},\qquad t=\int \frac{d\tau}{R(\tau)}
\label{EX6}
\end{equation} 
transforms the RW line element into 
\begin{equation}
ds^2=e^{-2\alpha(\tau,\rho)} \left[(1+\gamma_0 r)c^2dt^2-\frac{dr^2}{(1+\gamma_0 r)}-r^2d\Omega_2\right],
\label{EX7}
\end{equation}
where 
\begin{equation}
\gamma_0=2(-K)^{1/2},\qquad e^{-\alpha(\tau,\rho)}=R(\tau)\frac{(1-\gamma_0\rho/4)}{(1+\gamma_0\rho/4)}.
\label{EX8}
\end{equation}                                 
With a conformal transformation of the form $g_{\mu\nu}\rightarrow g_{\mu\nu}e^{2\alpha(\tau,\rho)}$ bringing the line element in equation (\ref{EX7}) to the Schwarzschild coordinate system, we see that in the rest frame of a galaxy that is comoving with the Hubble flow, conformal cosmology  acts like a universal linear potential. With conformal cosmology naturally leading  \cite{Mannheim2006} to a spatially open RW geometry with negative spatial curvature $K$, the coefficient $\gamma_0$ of the universal linear potential will be real, just as is needed to enable $\gamma_0 c^2 r/2$ is to serve as a gravitational potential. Since the measured rotational velocities in galaxies are non-relativistic we can incorporate the universal linear potential using weak gravity, with (\ref{E5}) then being augmented \cite{Mannheim1997} to 
\begin{equation}
v^2_{\rm TOT}(R)=v^2_{\rm LOC}({\rm disk})+\frac{\gamma_0c^2 R}{2}.
\label{EX9}
\end{equation}                                 

To determine the specific  contribution due to inhomogeneities in the cosmological background, we return to the third and fourth integrals in equation (\ref{EX4}) to which such cosmological inhomogeneities will contribute. With our interest being in applying  equation (\ref{EX4}) on galactic distance scales, we note that since such scales are much smaller than any typical cluster of galaxies scale $r_{\rm clus}$ that would be associated with cosmological inhomogeneities, for the purposes of evaluating $v^2_{\rm TOT}(R \ll r_{\rm clus})$ we can replace the lower limits in each of those two integrals by $r_{\rm clus}$. Thus, on setting
\begin{equation}
\kappa =\frac{1}{6}\int_{r_{\rm clus}}^{\infty} dr^{\prime}\,r^{\prime }f(r^{\prime}).
\label{EX10}
\end{equation}                                 
we augment equation (\ref{EX9}) to \cite{Mannheim2011e,Mannheim2012a}
\begin{equation}
v^2_{\rm TOT}(R)=v^2_{\rm LOC}({\rm disk})+\frac{\gamma_0 c^2R}{2}-\kappa c^2R^2,
\label{E20}
\end{equation}                                 
with associated asymptotic limit 
\begin{equation}
v_{{\rm TOT}}^2(R) \rightarrow \frac{N^*\beta^*c^2}{R}+
\frac{N^*\gamma^*c^2R}{2}+\frac{\gamma_0c^2R}{2}-\kappa c^2R^2.
\label{E21}
\end{equation} 

With $v^2_{\rm LOC}({\rm disk})$ readily generalizing (see equation (\ref{EX13}) below) to include more than one visible component (as well as an optical disk of stars galaxies can possess spherical stellar bulges and a disk of atomic gas), equation (\ref{E20}) thus gives the expectation of conformal gravity for galactic rotational velocities.  With the mass of the atomic gas being measurable, even with its inclusion, $v_{{\rm TOT}}^2(R)$ would still only contain one free parameter per galaxy, namely the total number of stars $N^*$, and is thus highly constrained. In the original study of \cite{Mannheim1997} equation (\ref{EX9}) was used to fit an 11 galaxy sample, and good fitting was found with parameters 
\begin{equation}
\gamma^*=5.42\times 10^{-41}~{\rm cm}^{-1},\qquad \gamma_0=3.06\times10^{-30}~{\rm cm}^{-1}.
\label{EX11}
\end{equation} 
At the time the fitting presented in \cite{Mannheim1997} was made, it was not appreciated that equation (\ref{EX9}) would need to be augmented by the quadratic $-\kappa c^2R^2$ term.  Basically, the data that were available at the time did not go out far enough in distance for any possible large distance behavior other than a linear growth in $v^2_{\rm TOT}(R)$ to matter. However as more data accumulated it became clear that continuing linear growth was not being found in the data. When a large data sample of 111 galaxies was explored \cite{Mannheim2011e,Mannheim2012a} a set of 21 of them was identified \cite{Mannheim2011e} in which the expectation of equations (\ref{EX9}) and (\ref{EX11}) completely overshot the data at large distances. Remarkably, it was found in \cite{Mannheim2011e,Mannheim2012a} that with the addition of just the one universal $-\kappa c^2R^2$ term with a repulsive overall sign and a magnitude 
\begin{equation}
\kappa=9.54\times 10^{-54}~{\rm cm}^{-2},
\label{EX12}
\end{equation} 
and with the continuing use of the values for $\gamma^*$ and $\gamma_0$ given in  (\ref{EX11}), fitting to the entire 21 galaxies could then be brought into accord with data, with fitting to none of the other  90 galaxies in the 111 galaxy sample being found to be adversely affected. The realization that the $\kappa$ term was significant on the largest available galactic distance scales underscores the value in having a data sample large enough that one could uncover unanticipated regularities. Subsequently, equations (\ref{E20}), (\ref{EX11}) and (\ref{EX12})  were successfully applied \cite{O'Brien2012} to a further set of 27 galaxies, and below we describe the fitting to the full 138 galaxy set in detail. In addition, we  present an application of our theory to three tidal dwarf galaxies. With only one free parameter per galaxy the successful fitting to a total of now 141 galaxies is very encouraging for the conformal  theory.

Self-consistently, we note that the value we obtain for $\gamma^*$ ensures that the linear potential of the sun is indeed negligible for the solar system, with the standard gravitational solar system phenomenology being left intact, just as desired. Additionally, we note that $\gamma^*$ is so small that one needs to go to a galactic system ($N^*$ of order $10^{10}$ or so, and $R$ of order a galactic distance scale) in order for the $N^*\gamma^*c^2R/2$ term to build up into something that could compete with Newtonian gravity. Moreover, the magnitude of the $\gamma_0$ parameter shows it to indeed be cosmological in scale, and with $\gamma_0/\gamma^*\sim 5\times 10^{10}$, the $\gamma_0$ term also first manifests itself on galactic scales. Finally, we note that with $\kappa^{-1/2} \sim 3 \times 10^{26}~{\rm cm}$ the parameter $\kappa$  is indeed a cosmological  inhomogeneity distance scale, again just as required.   Beyond being able to fit data at all with the very compact and highly constrained equation (\ref{E20}), we note that the specific values that we obtain for the $\gamma^*$, $\gamma_0$ and $\kappa$ parameters are completely compatible with the motivation that led us to consider them in the first place.

\section{Details of the Fitting Formalism}
\label{s2}
\medskip

In applying the conformal theory to galactic rotation curves we need to specify the local contribution to the rotational velocity due to each of the individual visible matter components within each galaxy. These components typically consist of a stellar disk, a stellar bulge, and atomic gas. Both of the stellar components are located within the optically visible disk, with any stellar bulge being located in the galactic center. For a stellar disk with scale length $R_0$ the optical disk typically becomes faint at a distance $R$ of order $4R_0$ or so. The atomic gas is distributed both within and beyond the optical disk. Since the Newtonian contribution of the stellar material (disk and bulge combined when applicable) would lead to a Keplerian fall off in rotational velocities in the $R>4R_0$ region, study of rotation curves via HI atomic gas radio observations provides the primary probe  of this expectation in the $R>4R_0$ region. With the measured circular velocities showing no sign of any Keplerian fall off beyond $R=4R_0$, and with the Newtonian contribution of the atomic gas itself not being substantial enough to affect the circular velocities in any significant way, the $R>4R_0$ HI radio data provide the clearest evidence that there is a missing mass problem in galaxies. To test whether the conformal theory can account for the detected departures from the luminous Newtonian expectation without any need to introduce dark matter,  we have therefore chosen to fit the HI rotation curve data of a large and very comprehensive set of 138 galaxies (we fit the HI data over the full available $R$ range and not just over the $R>4R_0$ range of course), with the data that we use being the best HI data available in the astronomical literature.

Of the galaxies we study, either the atomic gas mass is much less than the stellar mass (the case in high surface brightness (HSB) spirals) or the atomic gas mass and the stellar mass are both small (typically the case in low surface brightness (LSB) spirals or dwarf (DWF) spirals). With the atomic gas being distributed in a thin disk, then just like the stellar disk we shall model the gas profile as an exponential disk with an effective surface mass density $\Sigma_{\rm gas} (R)=\Sigma_0({\rm gas})e^{-R/R_0({\rm gas})}$. (Even though gas profiles are typically not as close to a single exponential disk as the stellar distributions, since the gas is not a substantial contributor to the total rotational velocity, our fits are not sensitive to our modeling of the gas profile.) Since the HI gas extends well beyond the optical disk, the effective scale length of the gas in any galaxy must be quite a bit larger than that of the disk stars in the same galaxy, and after modeling a few measured gas surface densities we found an effective gas scale length that was typically four times that of the accompanying stellar $R_0$. Since in every case we study there is little sensitivity  to the HI gas content, for each galaxy in our sample we shall use a gas scale length that is four times that of the stellar scale length of that galaxy. (For some of our chosen galaxies we had to use a model profile anyway since an actual gas profile had not been reported in the literature.) Since the atomic gas in a galaxy consists of both hydrogen and helium, we multiply the measured HI mass by 1.4 to account for primordial helium, to then obtain a total atomic gas mass $N_{\rm gas}M_{\odot}$ as expressed in solar mass units. With such a local gas distribution the contribution of the gas to rotation velocities  duplicates equation (\ref{E5}) and is given by 
\begin{eqnarray}
v^2_{\rm LOC}({\rm gas})&=&
\frac{N_{\rm gas}\beta^*c^2 R^2}{2R_0^3({\rm gas})}\left[I_0\left(\frac{R}{2R_0({\rm gas})}
\right)K_0\left(\frac{R}{2R_0({\rm gas})}\right)-
I_1\left(\frac{R}{2R_0({\rm gas})}\right)
K_1\left(\frac{R}{2R_0({\rm gas})}\right)\right] 
\nonumber\\
&+&\frac{N_{\rm gas}\gamma^* c^2R^2}{2R_0({\rm gas})}I_1\left(\frac{R}{2R_0({\rm gas})}\right)
K_1\left(\frac{R}{2R_0({\rm gas})}\right). 
\label{EG}
\end{eqnarray}

Some of the galaxies in our sample have a spherical bulge. With a spherical mass density $\sigma(r)$ per unit solar mass and a total mass  $N_{\rm bulge}M_{\odot}$ where $N_{\rm bulge}=4\pi \int  dr^{\prime}\,r^{\prime 2}\sigma(r^{\prime})$ the net gravitational potential due to a spherical bulge can then be obtained by integrating equation (\ref{E4}) over  the $\sigma(r)$ distribution, and is found to yield  circular velocities of the form \cite{Mannheim2006}
\begin{eqnarray}
v^2_{\rm LOC}({\rm bulge})&=&{4\pi\beta^* c^2\over r}\int_0^r
dr^{\prime}\,\sigma(r^{\prime}) r^{\prime 2}
\nonumber\\
&+&{2\pi\gamma^* c^2\over 3r}\int_0^r
dr^{\prime}\,\sigma(r^{\prime}) (3r^2r^{\prime 2}-r^{\prime 4})
+{4\pi\gamma^* c^2r^2\over 3}\int_r^{\infty} dr^{\prime}\,\sigma(r^{\prime})
r^{\prime}.
\label{A30}
\end{eqnarray} 
Finally, on including the gas and bulge contributions we augment equation (\ref{E20}) to
\begin{equation}
v^2_{\rm TOT}(R)=v^2_{\rm LOC}({\rm disk})+v^2_{\rm LOC}({\rm gas})+v^2_{\rm LOC}({\rm bulge})+\frac{\gamma_0 c^2R}{2}-\kappa c^2R^2,
\label{EX13}
\end{equation}                                 
with associated asymptotic limit 
\begin{equation}
v_{\rm TOT}^2(R) \rightarrow \frac{N_{\rm TOT}\beta^*c^2}{R}+
\frac{N_{\rm TOT}\gamma^*c^2R}{2}+\frac{\gamma_0c^2R}{2}-\kappa c^2R^2,
\label{EX14}
\end{equation} 
where  
\begin{equation}
N_{\rm TOT}=N^*+N_{\rm gas}+N_{\rm bulge}.
\label{EX15}
\end{equation} 

Equation (\ref{EX13}) is our main result, and we proceed below to apply it to galactic rotation curve data. As regards its use, we note that with the mass of the atomic gas $N_{\rm gas}M_{\odot}$ being measured directly  in the radio observations,  and with the parameters $\gamma^*$, $\gamma_0$ and $\kappa$ being universal (they can be determined by just a few rotation curve data points in just a single one of the 138 galaxies in our sample), in equation (\ref{EX13}) there is only one parameter that can vary from one galaxy to the next, namely the total stellar mass $(N^*+N_{\rm bulge})M_{\odot}$ in each galaxy. The galactic stellar mass is not measured directly since in photometric studies of the stars in the optical disk one instead measures the galactic luminosity. However, with  the conformal gravity contributions to $v_{\rm TOT}^2(R)$ growing with distance, in the inner regions of galaxies the Newtonian contribution is the most important, to thereby permit a determination of $(N^*+N_{\rm bulge})M_{\odot}$  from  inner region rotational velocity data alone. Encouragingly for our fits we find that when the values for  $(N^*+N_{\rm bulge})M_{\odot}$ that we obtain this way are divided by the measured luminosities we get ratios that are close to the mass to light ratio measured in the local solar neighborhood (i.e. within a factor of the mass to light ratio $M_{\odot}/L_{\odot}$ of the sun). Then, with $N^*+N_{\rm bulge}$ being determined by inner region data alone, the use of equation (\ref{EX13}) to fit the outer regions is effectively parameter free. The conformal gravity theory outer region predictions are thus highly constrained, and yet as we show, the theory is fully capable of accounting for the rotation curve systematics that are observed right across our 138 galaxy sample.

\section{Input Parameters for the Fits}
\label{s3}
\medskip

In order to apply equation (\ref{EX13}) to galaxies we need to supply input values for stellar scale lengths $R_0$ and distances $D$ to galaxies. In order to be able to test the conformal theory in as unbiased and standardized a way as possible,  we have established a common  baseline for selecting these parameters from data. Specifically, while there is often a range of  stellar disk scale lengths reported in the literature (a spread we took advantage of in a few cases),  by and large we have used the stellar scale length values as measured in the longest  wavelength band available for each galaxy (most often the K band). As regards distances to galaxies, we have used distances to galaxies as given by the NASA Extragalactic Database (NED). Determining such distances $D$ is a key requirement for our theory not just because such distances are needed to convert angles on the sky  to absolute scales for $R$, for $R_0$, and for the HI gas profile and total mass $M_{\rm HI}$,  but with the parameters $\gamma^*$, $\gamma_0$ and $\kappa$ being absolute quantities, any variation in $R$ would affect the absolute normalization of the contributions of the linear and quadratic terms in equation (\ref{EX13}). Our theory is thus very sensitive to distance determinations.

In the NED two procedures for determining distances are presented, one based on visual observations of nearby galaxies (usually Cepheid or Tully-Fisher determinations), and the other based on redshift measurements for more distant galaxies. The visually determined distance values are listed as a world mean value and its one standard deviation, while the redshift-based determinations depend on cosmological models and are listed with a spread in values depending on the model chosen. We have systematically used visually determined mean values and galactocentric cosmological model redshift-based distance determinations. With such an assumption we found \cite{Mannheim2011e,Mannheim2012a,O'Brien2012} that our fits were immediately falling on the data in all but 15 of the 138 cases, and in those 15 cases the fits were close enough that we could then bring them into agreement with data by allowing for the above noted uncertainties in the NED distance determinations. The fact that our fits work so well at the NED distances is thus a noteworthy achievement for our theory.

Also we note that there can be another uncertainty in the data, namely there can be some uncertainty in  the inclination angle of a galaxy.  Although this has no effect on the shape of the measured rotation curve, it would serve to change the normalization of the curve, since a change in the inclination angle would result in a  changed determination of the actual values of the rotational velocities.  Specifically, with the inclination angle being the angle between the normal to the plane of the galactic disk and our line of sight, a measured Doppler shift $v({\rm meas})$ is actually a measurement of the projection $v(i)\sin i$ at inclination angle $i$ along our line of sight (so that for edge-on galaxies $v(90^\circ)=v({\rm meas})$). At two different assumed inclinations the inferred velocities would thus be related as $v({\rm meas})=v(i_1)\sin i_1=v(i_2)\sin i_2$, with a decrease or increase in assumed inclination leading to an increase or decrease in inferred rotation velocity.  For five of the galaxies in our 138 galaxy sample we found it advantageous to allow for inclination uncertainties. Since we use single exponential models for both the stellar disk and the atomic gas profile (equivalent to a face-on $0^{\circ}$ inclination angle),  we have no need to take into consideration any inclination effects that would be needed if we were to use the actually measured stellar or gas profiles themselves.

\section{The Conformal Gravity Fits}
\label{s4}
\medskip

In making a selection of which galactic rotation curves to study, for each galaxy we required that there be both good optical photometry (as needed to determine the  explicit contribution due to the visible material in each galaxy), and reliable, well-accepted HI rotation curve data. For the HI data we explicitly sought data that extended the furthest in radial distance, so that we could explore the kinematic region that is the most sensitive to the linear and quadratic terms in equation (\ref{EX13}). Our sample contains galaxies that range in luminosity by as much as four orders of magnitude, ranging from very bright HSB galaxies (as bright as $2 \times10^{11}L^{\rm B}_{\odot}$), to LSB galaxies and to DWF galaxies (as dim as $3 \times 10^{7}L^{\rm B}_{\odot}$). In our sample there are 21 galaxies that go out to a last radial data point $R_{\rm last}$ that is more than $31~{\rm kpc}$, with the largest being Malin 1, a galaxy for which $R_{\rm last}$ is a mammoth $98~{\rm kpc}$.

Given these requirements, we have identified a sample consisting of 138 galaxies in total. The  sample is comprised of 18 galaxies from The HI Nearby Galactic Survey (THINGS), 30 galaxies from an Ursa Major study, 20 galaxies from an LSB galaxy study, 21 from a second LSB study, 22 miscellaneous galaxies, 20 galaxies from a dwarf galaxy study, 2 additional LSB galaxies, and  5 new THINGS dwarf galaxies. The miscellaneous 22 galaxy set contains many of the classic rotation curves that first established that there actually was a galactic missing mass problem. The galaxies we fit have frequently been studied in the literature for dark matter fitting, as well for as alternative gravity fitting based on  theories such as MOND \cite{Sanders2002,Swaters2010} or MSTG \cite{Brownstein2006}.

We reported conformal gravity fitting to the first  111 galaxies of our sample in \cite{Mannheim2012a} (with the largest 21 also being reported in \cite{Mannheim2011e}), and reported fitting to the remaining 27 in \cite{O'Brien2012}. To illustrate how our theory fares, for our purposes here we present  a few typical fits from each of the components of our sample.  From the THINGS sample we present fits  to NGC 3198, NGC 3521 and  NGC 5055 (all HSB). From the Ursa Major sample we present fits to NGC 3893, NGC 4183 (both HSB) and UGC 6923 (LSB). From the 20 galaxy LSB survey we present fits to F563-1, NGC 959 and UGC 128. From the 21 galaxy LSB survey we present fits to F579-V1, UGC 6614, and UGC 11557. From the 22 miscellaneous  galaxy set we present fits to NGC 247 (LSB) , NGC 2683 (HSB) and Malin 1 (HSB). From the 20 DWF  galaxy sample we present fits to UGC 731 and UGC 11707,  and from the 5 THINGS DWF galaxies we present a fit to UGC 5423. In Table 1 we present the input and output data associated with these selected galaxies and provide the relevant data sources for rotational velocities (column labelled $v$), luminosities ($L$), stellar disk scale lengths ($R_0$), and HI gas masses (${\rm HI}$). Of the 18 galaxies we present here two have spherical bulges, NGC 5055 and Malin 1. For theses two galaxies we find respective bulge masses $0.73\times 10^{10}M_{\odot}$ and $9.46\times 10^{10}M_{\odot}$. In Table 1 the mass to light ratios listed for these two galaxies are ratios of the total stellar mass (disk plus bulge) to total blue luminosity.

In Figure 1 we present the conformal gravity fitting to the rotational velocities (in ${\rm km}~{\rm sec}^{-1}$) together with their quoted errors as plotted as a function of radial distance (in ${\rm kpc}$). For each of the 18 selected galaxies we have exhibited the contribution due to the luminous Newtonian term alone (dashed curve), the contribution from the two linear terms alone (dot-dashed curve), the contribution from the two linear terms and the quadratic terms combined (dotted curve), with the full curve showing the total contribution. As we see, the tightly constrained equation (\ref{EX13}) captures the essence of the data, and does so without needing any dark matter whatsoever.  It is  a notable achievement for the theory that not only do we fit the data with only one free parameter per galaxy, viz. the galactic mass to light ratio, we are able to do so with fitted mass to light ratios that are typical of the mass to light ratio found in the local solar neighborhood.

As we can see from the fitting to the largest galaxies in Figure 1, were it not for the negative quadratic potential term the contribution of the linear terms would totally overshoot the data at the largest radial distances. Moreover, despite the fact that the quadratic term is universal and has no dependence at all on the galaxy-dependent $N_{\rm TOT}$ factor in (\ref{EX14}), nonetheless the quadratic term is able to arrest the linear rise in each and every case. So dramatic is this effect, that if we were to go to even larger radial distances, we would expect the rotation velocities to begin to decline and eventually be cut-off at some large radial distance of order $R\sim (\gamma_0+N_{\rm TOT}\gamma^*)/2\kappa$ where $v^2_{\rm TOT}(R)$ would otherwise become negative. In the conformal theory galaxies thus have some natural limiting size that is fixed by an interplay of the local galaxy with the global cosmologically generated $\gamma_0$ and $\kappa$ terms.  This eventual decline in the rotation curve provides a falsifiable test of the conformal gravity theory.  To illustrate this effect,  in Figure 2 we provide extended distance predictions for some six of the largest galaxies in our sample (NGC 3198, NGC 3521, NGC 5055, UGC 128, NGC 2683, and Malin 1).  With the last few data points in  UGC 128 and NGC 2683 perhaps already suggesting the beginnings  of a fall off, testing the prediction of an eventual fall off in rotation curves could  be within observational reach.  

In regard to the issue of a possible limiting size to galaxies, we take note of the study of \cite{Nandi2012}, in which it was noted that there would be constraints on circular orbits if the potential is treated as a one-body potential. Specifically, if we introduce a point particle Lagrangian $L=\dot{r}^2/2+r^2\dot{\phi}^2/2 -V(r)$ for motion in a plane with central potential $V(r)$, we obtain circular orbits that obey $\ddot{r}=r\dot{\phi}^2-V^{\prime}(r)=0$ and $r^2\dot{\phi}=J$, and thus $J^2/r^3=V^{\prime}(r)$. If we define an effective potential of the form $V_{\rm eff}(r)=V(r)+J^2/2r^2$ (so that $\ddot{r}=-V_{\rm eff}^{\prime}(r)$), then the radii of circular orbits have to obey $V^{\prime}(r)\geq 0$ if $r\dot{\phi}$ is to be real, and have to obey $V^{\prime\prime}_{\rm eff}=V^{\prime\prime}(r)+3V^{\prime}(r)/r \geq 0$ if the orbit is to be stable. For a potential of the form $V(r)=\gamma c^2 r/2-\kappa c^2 r^2/2$, $V^{\prime}(r)$ vanishes at $r=\gamma/2\kappa$, while the quantity $V^{\prime\prime}(r)+3V^{\prime}(r)/r $ vanishes at $r=3\gamma/8\kappa$. Thus only orbits with  $r\leq 3\gamma/8\kappa$ would be stable, and those with $3\gamma/8\kappa\leq r \leq \gamma/2\kappa$ would not be. Even with this stability condition rotation velocities would still undergo a fall off, but the radius of the largest stable orbit would be smaller than that associated with the vanishing of $v^2_{\rm TOT}(R)$. 

While the above stability analysis would apply if the problem were a one-body problem, in the conformal theory the potential itself arises through an interplay between each local galaxy and the global cosmology, and is thus actually many body in nature. Moreover, the very existence of galaxies at all requires that inhomogeneities form as perturbations in an otherwise uniform cosmological background. With these perturbations generating the $\kappa$ term  (c.f.  equation (\ref{EX10})), galaxies generated this way will already be stable as long as the cosmological perturbations themselves are stable. At the present time a theory for the growth of conformal gravity fluctuations is still being developed \cite{Mannheim2012b}, and a cosmological stability analysis has yet to be made. It could thus be very instructive to explore both theoretically and observationally where (and of course whether) galaxies might actually terminate.

In our study we have found that the simple universal formula given in equation (\ref{EX13}) accounts for a large body of data. The reason that this formula is able to do so is because there is a lot of universality in the data themselves. Specifically, across the entire 138 galaxies in our sample, and as specifically noted in Table 1 for our 18 selected galaxies, the value of the centripetal acceleration $(v^2/c^2R)_{\rm last}$ at the last data point is found to be close in magnitude to the numerical value we found for the universal $\gamma_0$  parameter. It is this universality that drives the fits,  with such universality naturally being present in the  conformal theory because $\gamma_0=2(-K)^{1/2}$ has a natural origin in the spatial curvature of the universe. At the present time there is no indication of the emergence of any such universality for $(v^2/c^2R)_{\rm last}$ in dark matter theory, and to account for the 138 galaxies using dark matter no less than 276 additional adjustable free parameters are required (two free numerical parameters per dark matter halo). Since our equation (\ref{EX13}) does account for the data, if it is to be dark matter theory that is to be the correct explanation of the missing mass problem, then one would need to be able to derive equation (\ref{EX13}) in it. Given the phenomenological success of equation (\ref{EX13}), and given the fact that it is derived from a fundamental gravitational theory, the conformal gravity theory poses a quite considerable challenge to the standard dark matter paradigm.

\section{Tidal Galaxies}
\label{s5}
\medskip

We conclude this review with an application of the conformal gravity theory to tidal dwarf galaxies (TDGs). These dwarf galaxies are formed in the tidal tails of collisions of disk galaxies, and are thought \cite{Barnes1992} to be predominantly composed of material expelled from the galactic disk of a parent galaxy. With any dark matter present in the parent galaxies expected to predominantly be in spherical haloes, tidal galaxies should thus have a very low dark matter content, and thus should not themselves be expected to possess the substantial spherical dark matter haloes that are ordinarily required to accompany and stabilize disk galaxies in standard gravity.  In consequence, in the standard dark matter picture TDG  rotation curves should not be expected to display any substantial mass discrepancies.  However, in any explanation of the missing mass problem that involves a change in Newton's law of gravity on galactic distance scales as opposed to a change in the matter content of galaxies, one should expect mass discrepancies. Tidal dwarf galaxies thus provide a quite unusual laboratory for exploring the missing mass problem. Rotation curve data have become available for three TDGs associated with the parent  galaxy NGC 5291 \cite{Bournaud2007} (see especially the accompanying supporting online material), and it has been shown \cite{Bournaud2007} that there are in fact mass discrepancies, and that a good accounting of the data can be provided by MOND \cite{Gentile2007}.  Here we show that conformal gravity can also provide a good accounting of the data.

Because of their locations the three TDGs in NGC 5291 are designated NGC 5291N, NGC 5291S, and NGC 5291SW. HI rotation curves are available for all three, and for NGC 5291N some inner region optical rotation curve data are available as well \cite{Bournaud2004}. For the TDGs there is little reported photometry in the literature and for only one, NGC 5291N, is there a reported luminosity. However, there are good HI gas profiles for all three TDGs, and so we can extract effective gas scale lengths. From them we can then infer effective stellar scale lengths by dividing the gas scale lengths by a factor of four.  With the HI gas masses being known, we shall estimate the luminosities of NGC 5291S and NGC 5291SW by assuming that the luminosity scales uniformly with the total gas mass for all three TDGs. With the luminosity of NGC 5291N being reported in the V band, to convert to a blue luminosity we shall use the $B-V=0.3$ modulus that was found \cite{Duc1998} to be characteristic of the debris around NGC 5291. Also, for the total gas mass  $M_{\rm gas}^{\rm tot}$ in each TDG we use the combined atomic hydrogen, atomic helium and molecular hydrogen total mass values reported in \cite{Bournaud2007}. With there only being redshift determinations of the distance $D$ to NGC 5291N and nothing at all reported in the NED for NGC 5291S or NGC 5291SW, we shall follow \cite{Bournaud2007} and \cite{Gentile2007} and for all three TDGs use the mean redshift determined distance of 62 Mpc  for NGC 5291N as given in the NED.  Finally, again following  \cite{Bournaud2007} we initially take the inclination angle $i$ of each of the three TDGs to be $45^{\circ}$. In our fitting we shall follow \cite{Mannheim2012a} and \cite{O'Brien2012} and require the mass to light ratio of each galaxy to be at least $0.2$.

Using equation (\ref{EX13}) we obtain the plots given in  Figure 3. As can be seen, the conformal theory captures the essence of the TDG data. Since there is a reported uncertainty  in inclination angle for NGC 5291N \cite{Bournaud2007,Gentile2007}, for it we can use inclination angles up to $55^{\circ}$. At $i=55^{\circ}$ we obtain the improved fit to NGC 5291N given in Figure 4. We list all of the input and output parameters for our fits in Table 2.

As we see the conformal theory nicely accounts for the tidal dwarf galaxy data. That it is able to do so is because, as can be seen in Table 2, just as with our earlier 138 galaxies, for all three TDGs the value of the centripetal acceleration at the last data point is again found to be very close to the numerical value obtained for $\gamma_0$. In conclusion therefore, we note that given its success in fitting no less than 141 galactic rotation curves with only one free parameter per galaxy,  and given in addition its theoretical underpinnings, conformal gravity can realistically be advanced as a candidate alternate theory of gravity.

\ack
\medskip

The authors wish to thank Dr.~J.~R.~Brownstein, Dr.~W.~J.~G.~ de Blok, Dr.~J.~W.~Moffat, Dr.~ S.~S.~McGaugh, Dr.~R.~A.~Swaters, Dr.~S.-H.~Oh, Dr. ~B. ~Famaey, and Dr. F.~Bournaud for helpful communications, and especially for providing their galactic data bases. This research has made use of the NASA/IPAC Extragalactic Database (NED) which is operated by the Jet Propulsion Laboratory, California Institute of Technology, under contract with the National Aeronautics and Space Administration. This review is based in part on a talk given by one of of us (JGO) at the 8th Biennial Conference on Classical and Quantum Relativistic Dynamics of Particles and Fields held in Florence, Italy in May 2012. JGO would like to thank the International Association of Relativistic Dynamics  for the kind hospitality of the conference and would like to thank the Istituto Nazionale di Fisica
Nucleare for related financial support.

\begin{table}[H]
\begin{center}
\caption{Properties of the Selected 18 Galaxy Sample}
\medskip
\scriptsize
\begin{tabular}{l c c c c c c c c c c} 
\hline\hline
 \phantom{00}Galaxy  $\!$&$D \!$& $L_{\rm B}\!$ & $(R_0)_{\rm disk}\!$  & $R_{\rm last} \!$ &  $M_{\rm HI} \!$ & $M_{\rm disk}\!$ &  $ 
(M/L) _{\rm stars}\!$ & $(v^2 / c^2 R)_{\rm last}\!$ & Data~Sources $\!$\\  
&  $({\rm Mpc})  \!$&  $(10^{10}L^{\rm B}_{\odot})\!$&$({\rm kpc}) \!$& $({\rm kpc}) \!$& {$(10^{10} M_\odot)\!$} & {$(10^{10}
M_\odot)\!$} & ({$M_{\odot}/L^{\rm B}_{\odot}\!$}) & {$(10^{-30}~\texttt{cm}^{-1})\!$} & $v~~~~L~~~R_0~~{\rm HI}\!$
\\
\hline
NGC \phantom{0}3198 & \phantom{0}14.1 &   3.24   & \phantom{0}4.0 & 38.6 & 1.06 & 3.64 & \phantom{0}1.12& 2.09&\cite{deBlok2008} \cite{Walter2008} \cite{Wevers1986} \cite{Walter2008}  
\\
NGC \phantom{0}3521 & \phantom{0}12.2&   4.77  & \phantom{0}3.3 & 35.3 & 1.03 & 9.25 &  \phantom{0}1.94
&4.21&\cite{deBlok2008} \cite{Walter2008} \cite{Leroy2008} \cite{Walter2008}   
\\
NGC \phantom{0}5055 & \phantom{0}\phantom{0}9.2&  3.62    &\phantom{0}2.9 & 44.4 &0.76& 6.04 &  \phantom{0}1.87& 
2.36&\cite{deBlok2008} \cite{Walter2008} \cite{Leroy2008} \cite{Walter2008}   
\\
NGC \phantom{0}3893 & \phantom{0}18.1 &   2.93   & \phantom{0}2.4 & 20.5 & 0.59& 5.00 &  \phantom{0}1.71 &
3.85 &   \cite{Verheijen2001} \cite{Sanders1998} \cite{Tully1996} \cite{Sanders1998}    \\
NGC \phantom{0}4183 & \phantom{0}16.7 &    1.04 & \phantom{0}2.9 & 19.5 & 0.30 &1.43 &  \phantom{0}1.38 &
2.36 &    \cite{Verheijen2001} \cite{Sanders1998} \cite{Tully1996} \cite{Sanders1998}   \\
UGC \phantom{0}6923& \phantom{0}18.0 &  0.30    & \phantom{0}1.5 & \phantom{0}5.3 &0.08 & 0.35 &  \phantom{0}1.18 &
4.43 &    \cite{Verheijen2001} \cite{Sanders1998} \cite{Tully1997} \cite{Sanders1998}  
\\
F563-1  &\phantom{0}46.8 &   0.14  &\phantom{0}2.9 & 18.2 & 0.29 & 1.35 &  \phantom{0}9.65 & 2.44 &  \cite{McGaugh2001} \cite{deBlok1997} \cite{deBlok1997} \cite{deBlok1996}     
\\
NGC \phantom{00}959 & \phantom{0}13.5 & 0.33    & \phantom{0}1.3& \phantom{0}2.9 &   0.05 & 0.37 & \phantom{0}1.11&7.43 &   \cite{Kuzio2008}  \cite{Fisher1981} \cite{Esipov1991}   \cite{Fisher1981} 
\\
UGC \phantom{00}128 &  \phantom{0}64.6 &    0.60  &  \phantom{0}6.9 & 54.8 & 0.73 & 2.75 &  \phantom{0}4.60 & 1.03 &  \cite{Verheijen1999} \cite{deBlok1997} \cite{vanderHulst1993} \cite{vanderHulst1993}     
\\
F579-V1 & \phantom{0}86.9&  0.56   & \phantom{0}5.2 & 14.7 & 0.21 & 3.33 &
\phantom{0}5.98 &  3.18 & \cite{McGaugh2001} \cite{deBlok1996} \cite{deBlok1997} \cite{deBlok1996}      
 \\
UGC \phantom{0}6614 &\phantom{0}86.2 &    2.11 & \phantom{0}8.2 & 62.7 & 2.07 & 9.70 & \phantom{0}4.60& 2.39 &\cite{McGaugh2001} \cite{deBlok2001} \cite{vanderHulst1993} \cite{vanderHulst1993} 
\\
UGC 11557 & \phantom{0}23.7 &  1.81   &  \phantom{0}3.0 & \phantom{0}6.7 &0.25 &0.37& \phantom{0}0.20& 3.49 &  \cite{McGaugh2001} \cite{deBlok2001} \cite{Swaters2002b} \cite{Swaters2002b}    
\\
NGC  \phantom{00}247 & \phantom{00}3.6 & 0.51    & \phantom{0}4.2 & 14.3 & 0.16 &
1.25 &  \phantom{0}2.43 & 2.94& \cite{Carignan1990} \cite{Carignan1985a} \cite{Carignan1985a} \cite{Carignan1990}
\\
NGC \phantom{0}2683 & \phantom{0}10.2 &   1.88   & \phantom{0}2.4 & 36.0 &0.15 & 6.03 &  \phantom{0}3.20
& 2.28 & \cite{Casertano1991} \cite{Casertano1991} \cite{Kent1985} \cite{Sanders1996}
\\
Malin~1 & 338.5 &   7.91   & 84.2 & 98.0 &5.40 &1.00 &  \phantom{0}1.32 & 1.77& \cite{Lelli2010} \cite{Lelli2010} \cite{Pickering1997} \cite{Lelli2010}
\\
UGC \phantom{00}731&\phantom{0}11.8& 0.07   & \phantom{0}2.4 & 10.3 & 0.16 & 0.32 &  \phantom{0}4.63 &
1.91&\cite{Swaters2003} \cite{Swaters2002a} \cite{Swaters2010} \cite{Swaters2002b}  \\ 
UGC 11707&\phantom{0}21.5& 0.11  & \phantom{0}5.8 & 20.3 & 0.68 & 0.99&  \phantom{0}8.76 &
1.77 &\cite{Swaters2003} \cite{Swaters2002a} \cite{Swaters2010} \cite{Swaters2002b} \\
UGC \phantom{0}5423 &\phantom{00}7.1& 0.01   & \phantom{0}0.6 & \phantom{0}2.0& 0.01 & 0.03 &  \phantom{0}2.01 &
1.82&\cite{Oh2011} \cite{Walter2008} \cite{Oh2011} \cite{Walter2008} \\ 
\hline
\end{tabular}
\label{table:tab1}
\end{center}
\end{table}

\bigskip

\begin{table}[H]
\begin{center}
\caption{Properties of the 3 Tidal Dwarf  Galaxies}
\medskip
\scriptsize
\begin{tabular}{l c c c c c c c c c} 
\hline\hline
\phantom{00}Galaxy\phantom{0}&\phantom{00}D  & $L_{\rm  B}$ & ~~$i$~~& $(R_0)_{\rm gas}$  & $R_{\rm last} $ &  $M_{\rm gas}^{\rm tot} $ & $M_{\rm disk}$ &  $ 
(M/L) _{\rm disk}$ & $(v^2 / c^2 R)_{\rm last}$  \\  
&   (Mpc)  &  $(10^{8}{\rm L}_{\odot}^{\rm B})$  & $~~^{\circ}~~$ &  (kpc) & (kpc) & {$(10^{8} M_\odot)$} & {$(10^{8}
M_\odot)$} & ({$M_{\odot}/L_{\odot}^{\rm B}$}) & {$(10^{-30}\texttt{cm}^{-1})$} \\
\hline
NGC 5291N &62.0& \phantom{0}9.5  & 45& 0.8 &  4.7  &7.7  & 3.4 & 0.35 &
5.7  \\
NGC 5291S &62.0& 10.7  & 45&  0.8&  5.2     &8.6  &2.1 & 0.20 &
1.9\\
NGC 5291SW &62.0& \phantom{0}5.7 &45  &0.8 &  2.7   &4.6 &1.1 &0.20&
3.4\\
NGC 5291N &62.0& \phantom{0}9.5 &55&0.8 & 4.7    &7.7 & 1.9 &0.20 &
4.3\\
\hline
\end{tabular}
\label{table:tab2}
\end{center}
\end{table}

\bigskip

\begin{figure}[H]
\epsfig{file=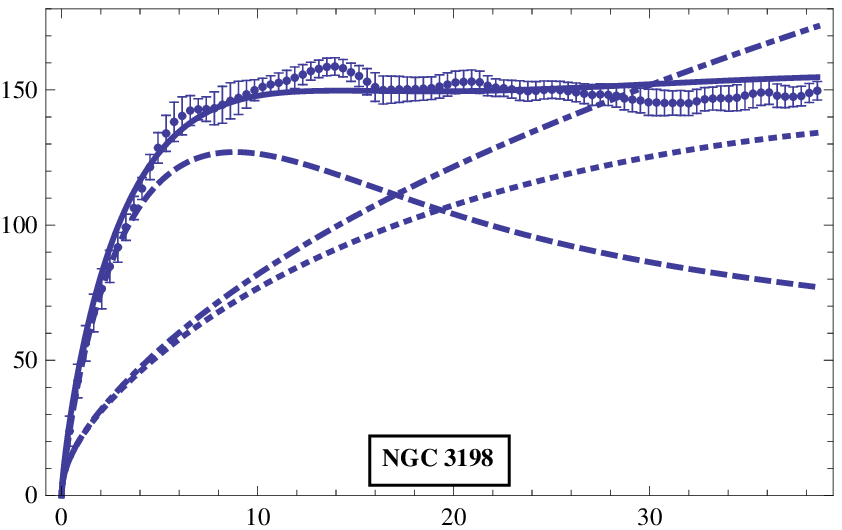,width=2.11in,height=1.2in}~~~
\epsfig{file=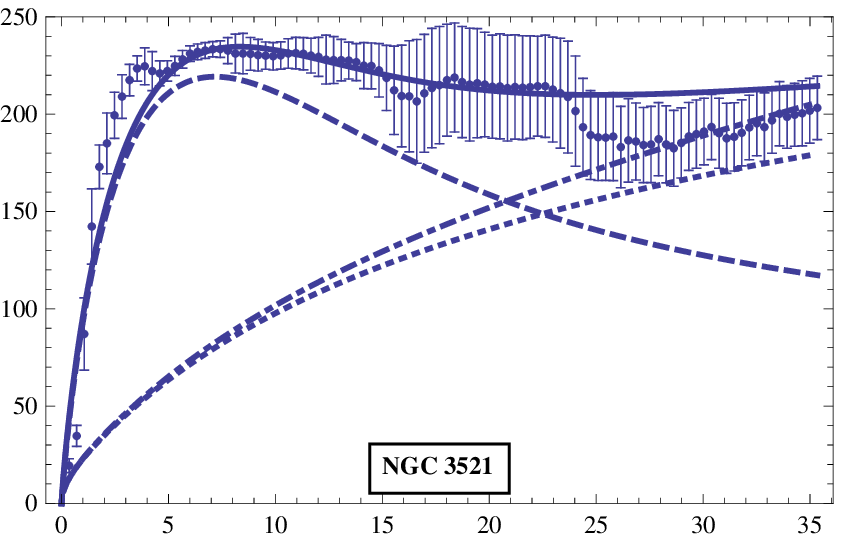,width=2.11in,height=1.2in}~~~
\epsfig{file=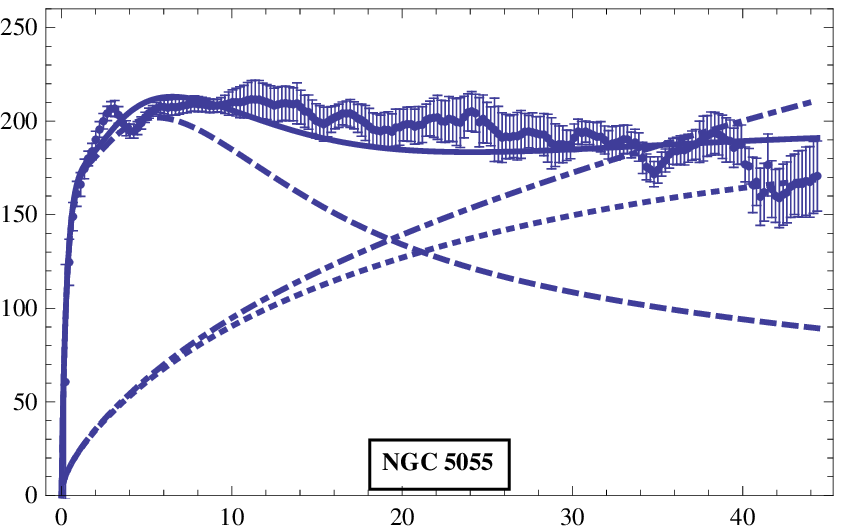,width=2.11in,height=1.2in}\\
\smallskip
\epsfig{file=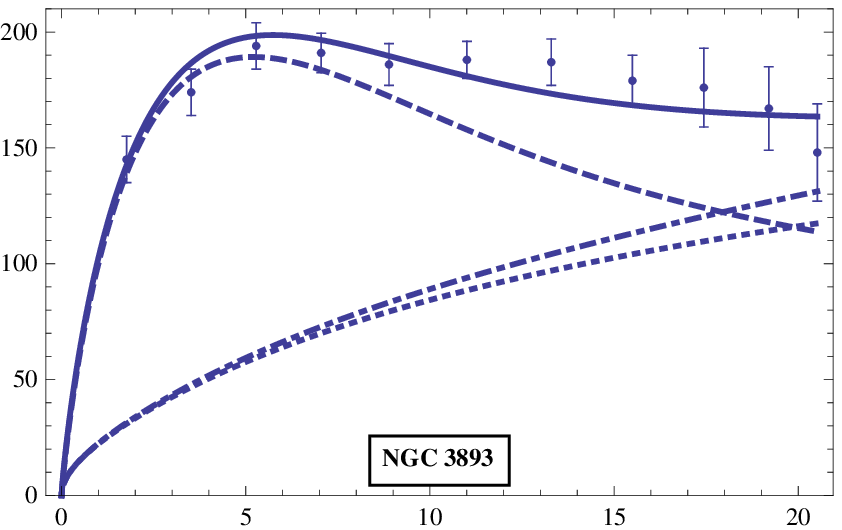,width=2.11in,height=1.2in}~~~
\epsfig{file=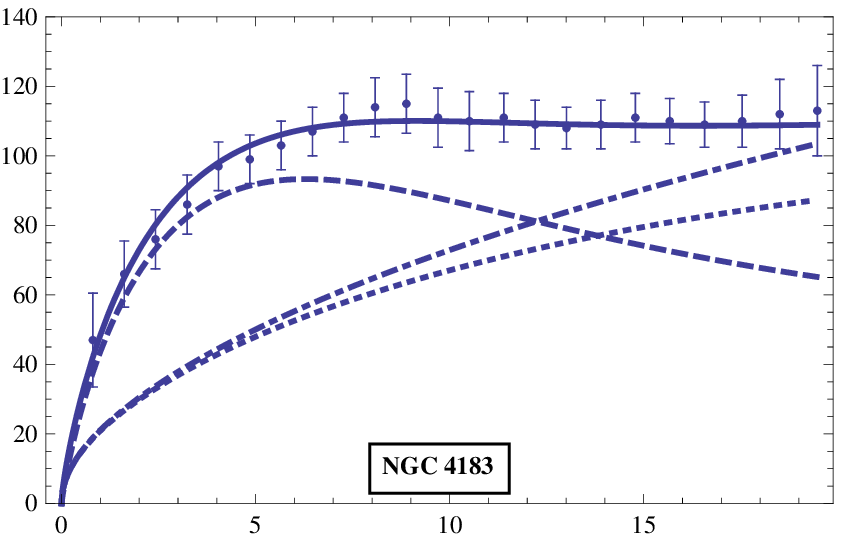,width=2.11in,height=1.2in}~~~
\epsfig{file=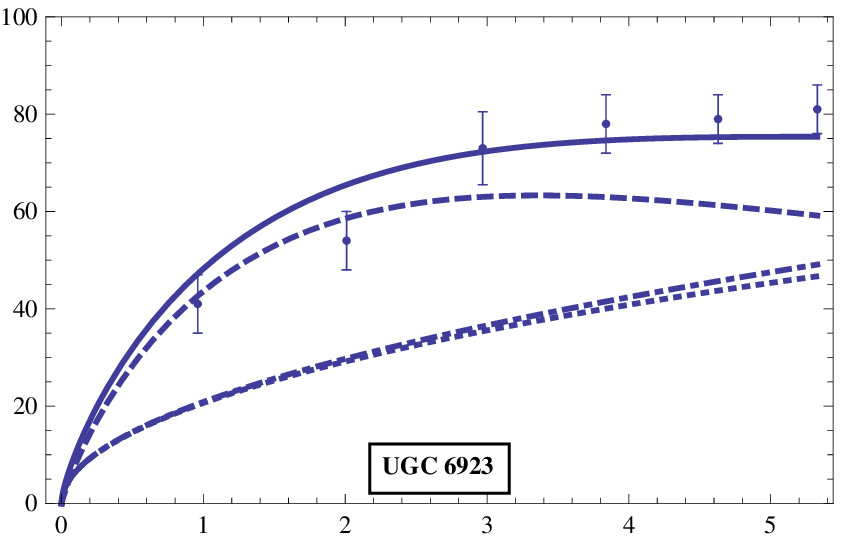, width=2.11in,height=1.2in}\\
\smallskip
\epsfig{file=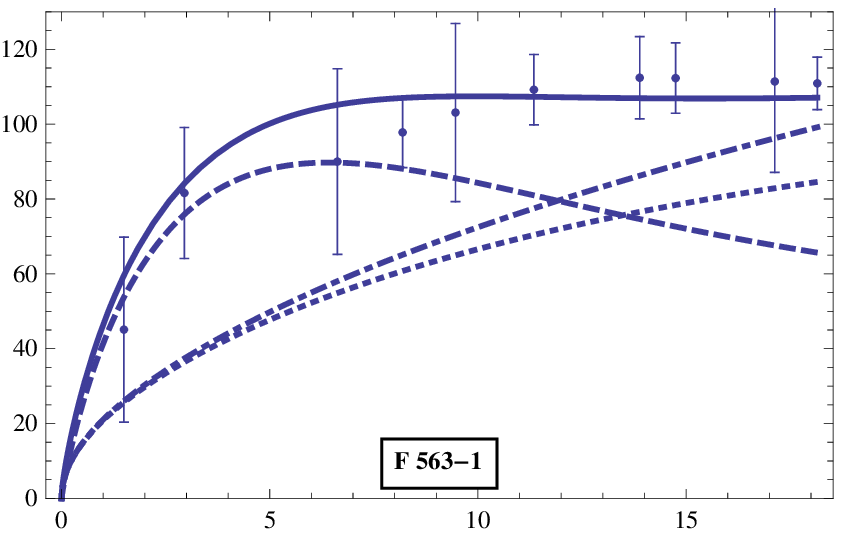,width=2.11in,height=1.2in}~~~
\epsfig{file=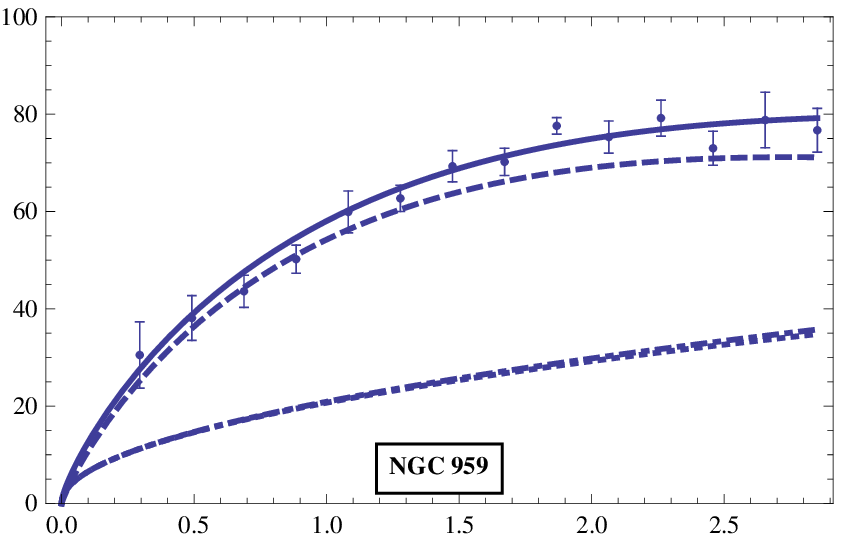,width=2.11in,height=1.2in}~~~
\epsfig{file=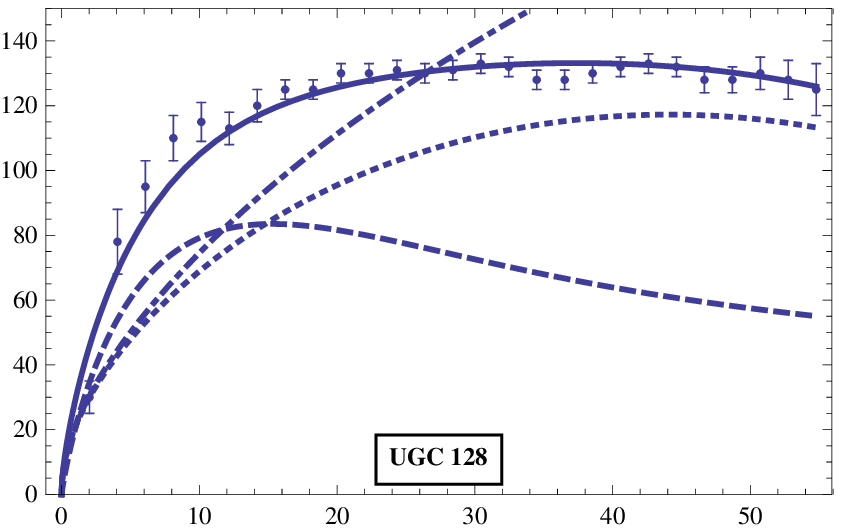,width=2.11in,height=1.2in}\\
\smallskip
\epsfig{file=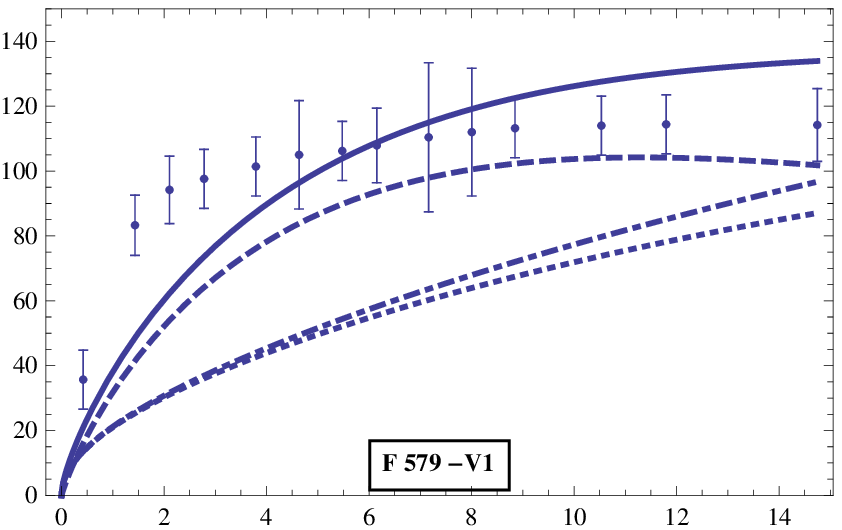,width=2.11in,height=1.2in}~~~
\epsfig{file=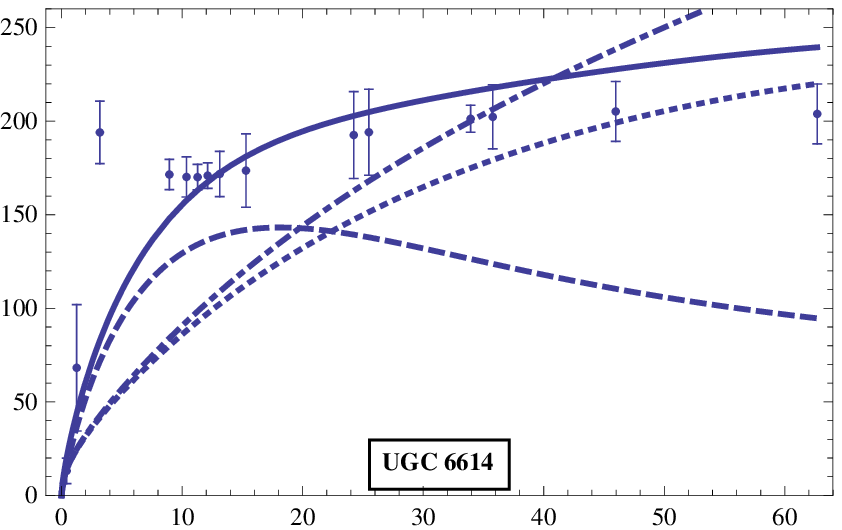,width=2.11in,height=1.2in}~~~
\epsfig{file=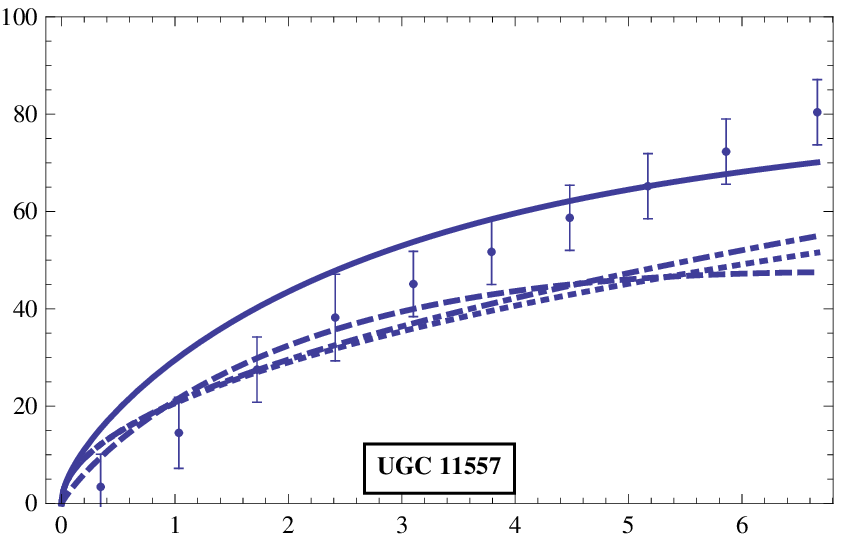,width=2.11in,height=1.2in}\\
\smallskip
\epsfig{file=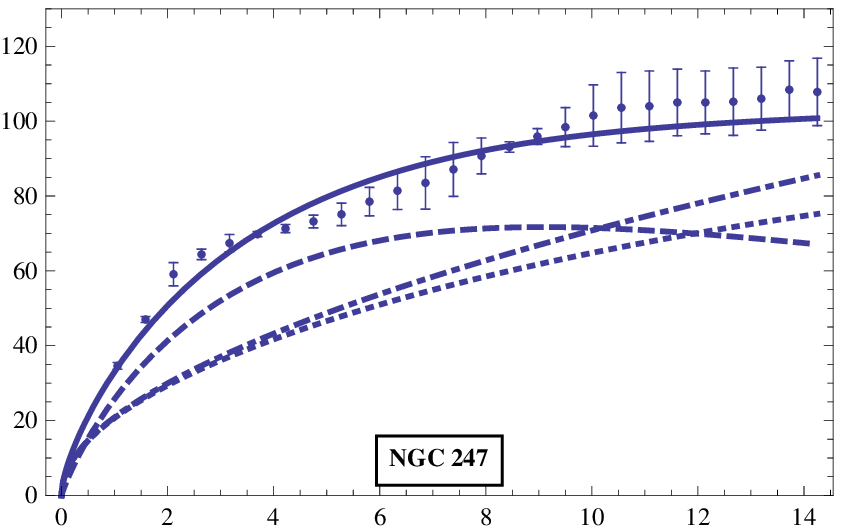,width=2.11in,height=1.2in}~~~
\epsfig{file=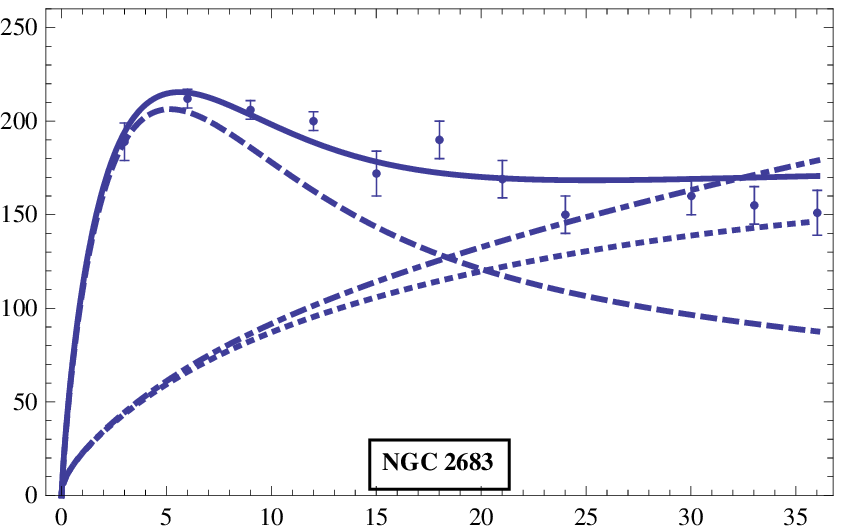,width=2.11in,height=1.2in}~~~
\epsfig{file=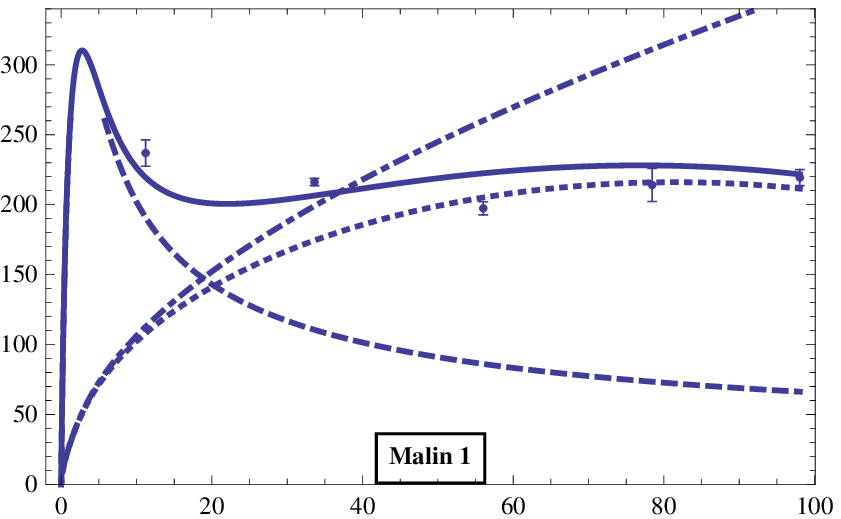,width=2.11in,height=1.2in}\\
\smallskip
\epsfig{file=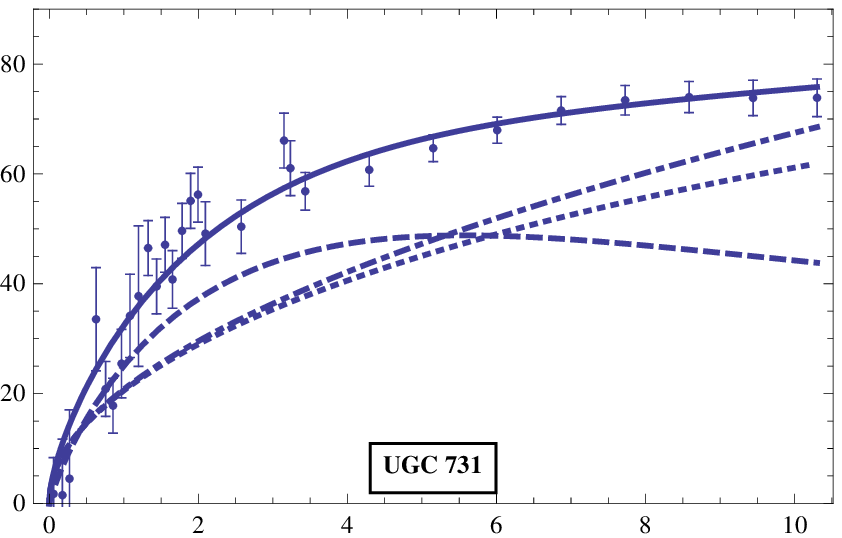,width=2.11in,height=1.2in}~~~
\epsfig{file=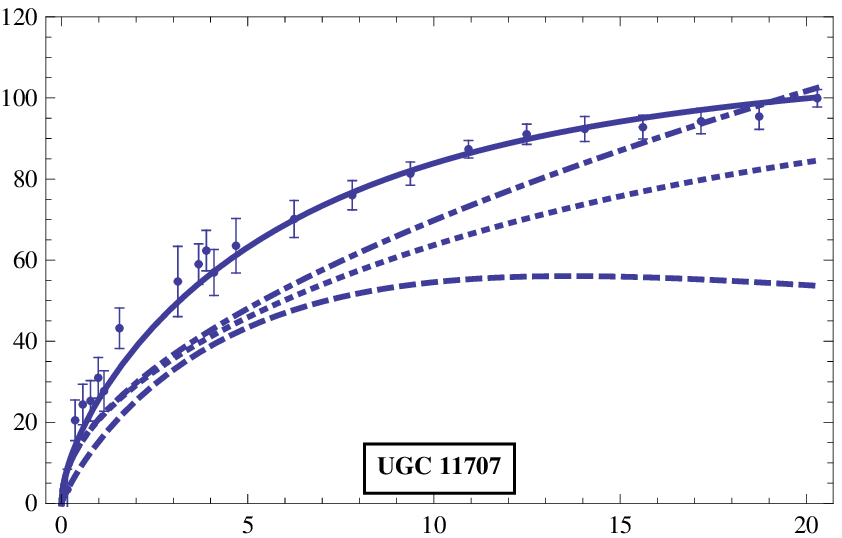,width=2.11in,height=1.2in}~~~
\epsfig{file=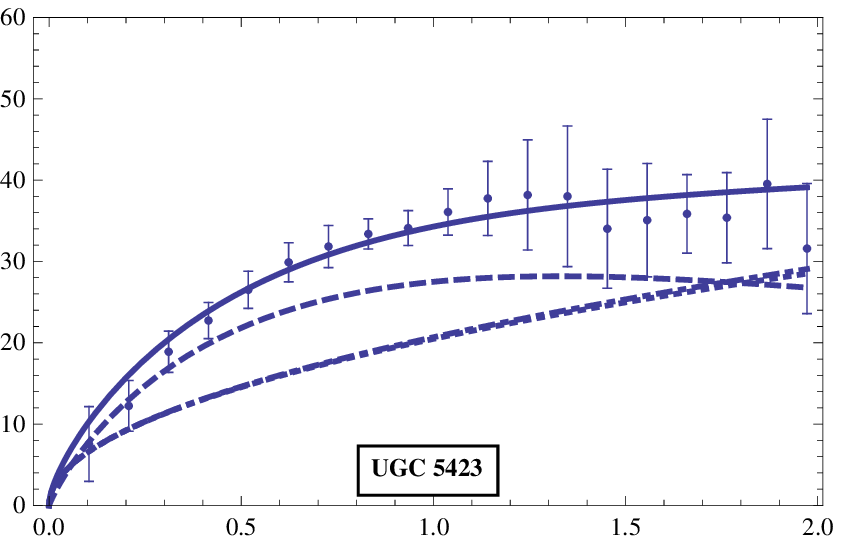,width=2.11in,height=1.2in}\\
\caption{Conformal gravity fitting to the rotational velocities (in ${\rm km}~{\rm sec}^{-1}$) of  the selected 18 galaxy sample with their quoted errors as plotted as a function of radial distance (in ${\rm kpc}$). For each galaxy we have exhibited the contribution due to the luminous Newtonian term alone (dashed curve), the contribution from the two linear terms alone (dot-dashed curve), the contribution from the two linear terms and the quadratic terms combined (dotted curve), with the full curve showing the total contribution. No dark matter is assumed.}
\label{Fig. 1}
\end{figure}

\begin{figure}[H]
\epsfig{file=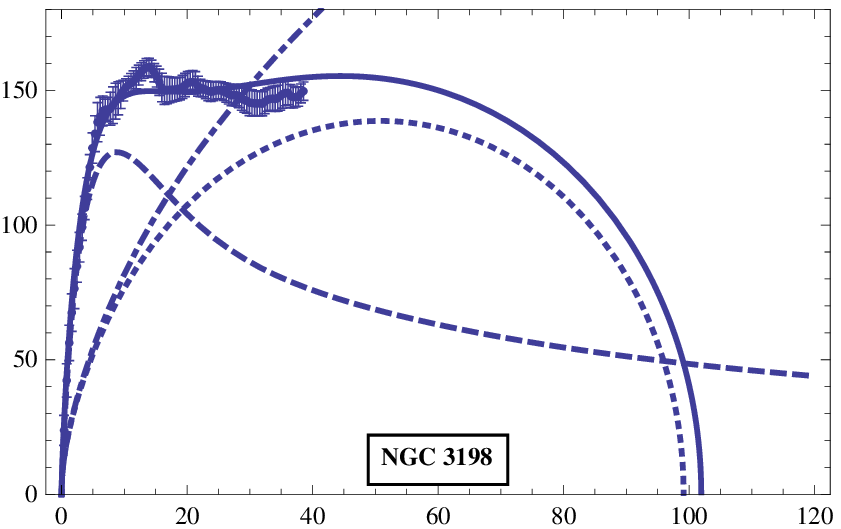, width=2.23in,height=1.2in}~~~
\epsfig{file=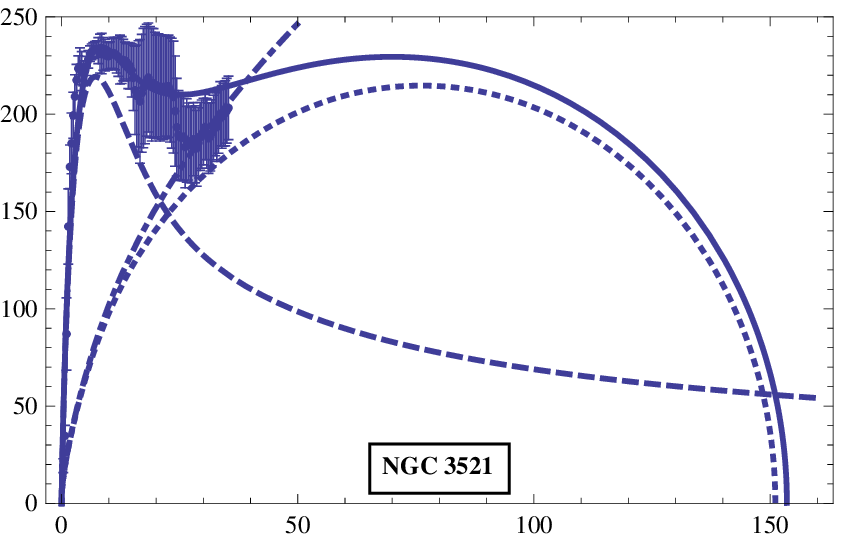, width=2.23in,height=1.2in}~~~
\epsfig{file=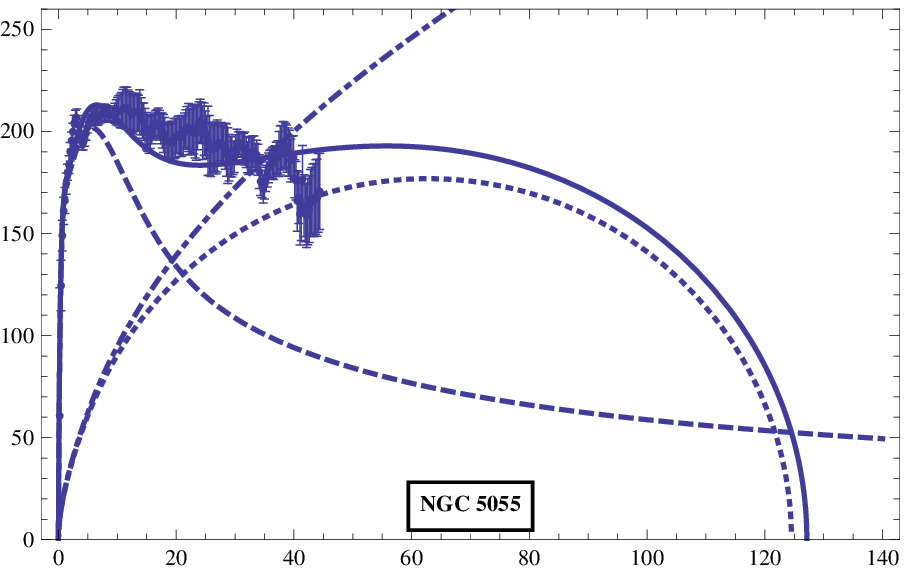, width=2.23in,height=1.2in}\\
\smallskip
\epsfig{file=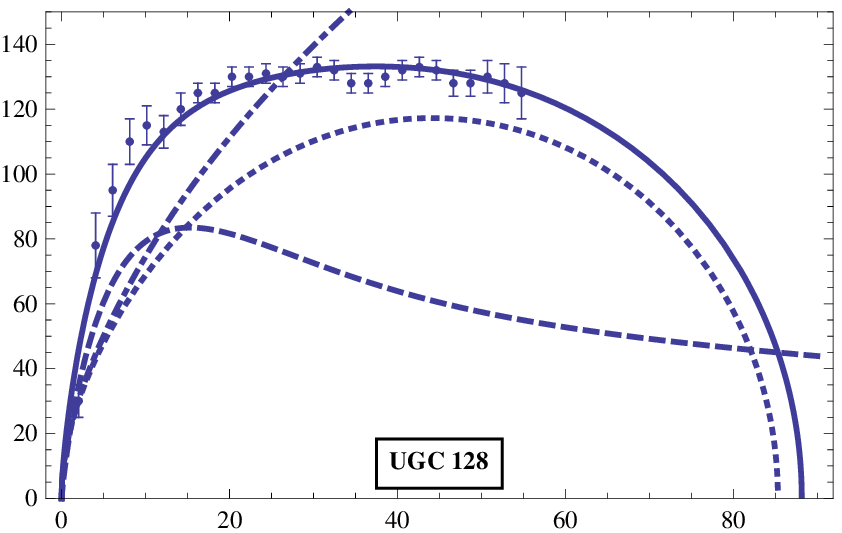, width=2.23in,height=1.2in}~~~
\epsfig{file=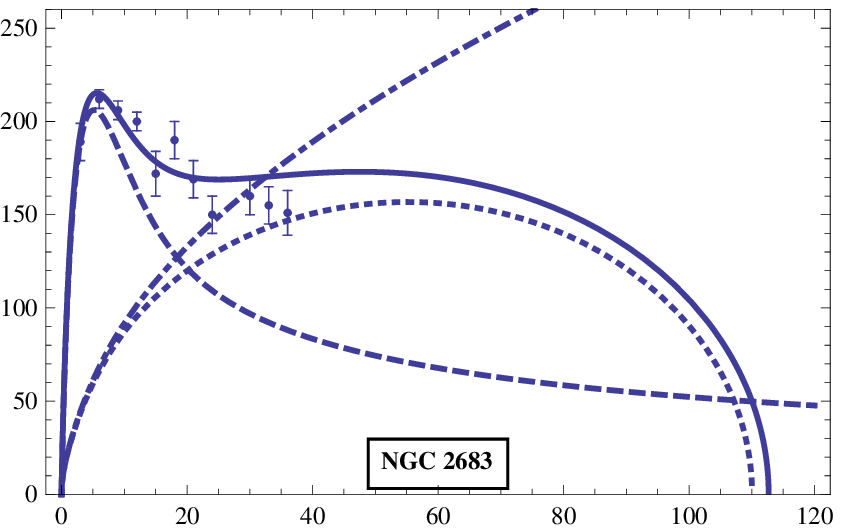, width=2.23in,height=1.2in}~~~
\epsfig{file=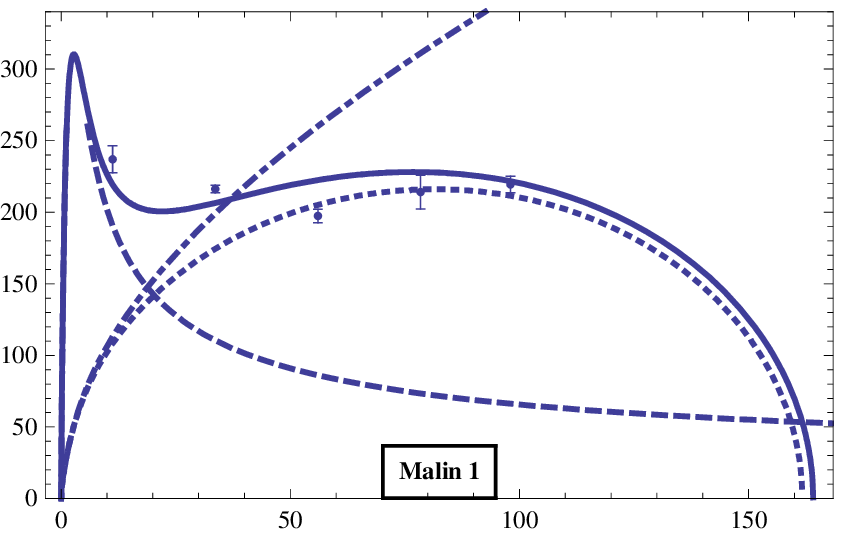, width=2.23in,height=1.2in}\\
\caption{Extended distance predictions for NGC 3198, NGC 3521, NGC 5055, UGC 128, NGC 2683 and Malin 1.  The curves are the same as in Figure 1.}
\label{Fig. 2}
\end{figure}

\begin{figure}[H]
\epsfig{file=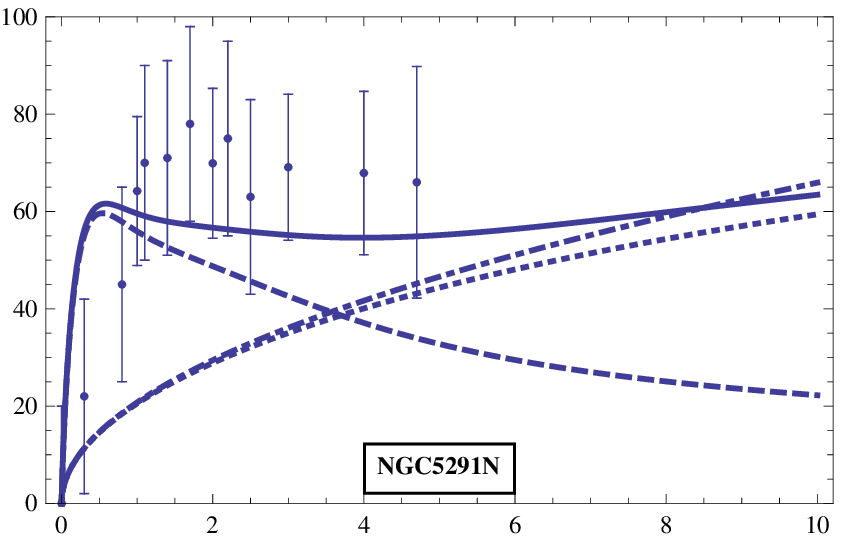, width=2.23in,height=1.2in}~~~
\epsfig{file=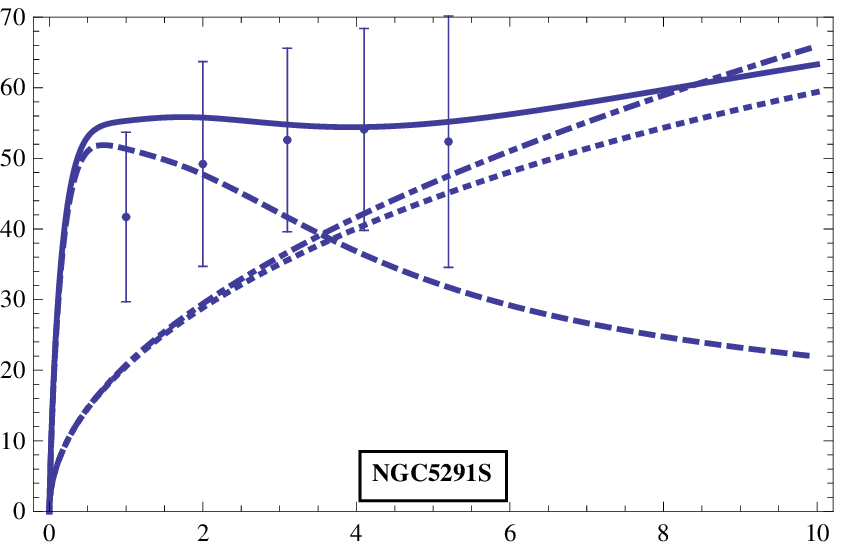, width=2.23in,height=1.2in}~~~
\epsfig{file=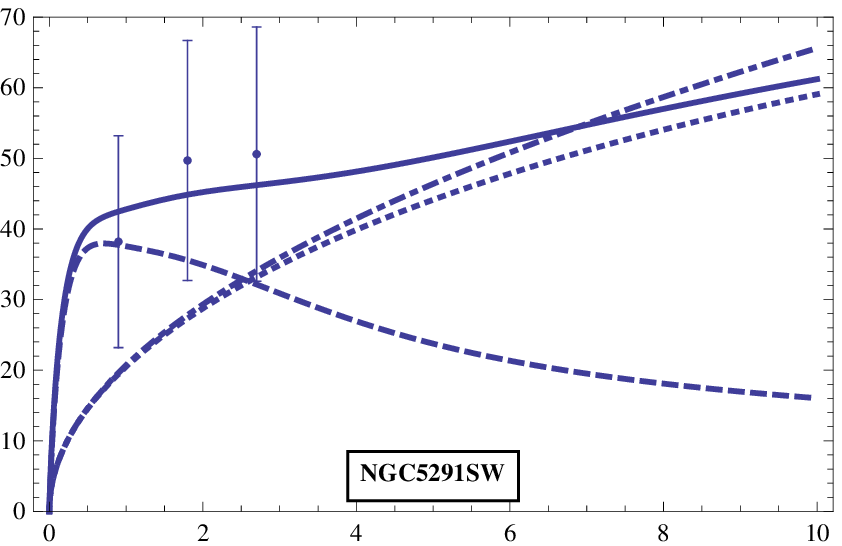, width=2.23in,height=1.2in}\\
\caption{Rotation curves for the tidal dwarf galaxies NGC 5291N, NGC 5291S, and NGC 5291SW, all with inclination $45^{\circ}$.  The curves are the same as in Figure 1.}
\label{Fig. 3}
\end{figure}

\begin{figure}[H]
\epsfig{file=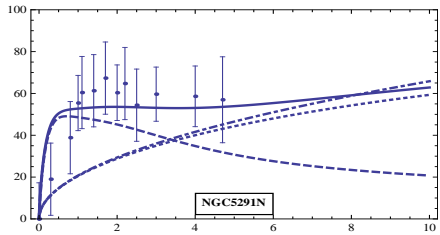, width=2.23in,height=1.2in}\\
\caption{Rotation curve for the tidal dwarf galaxy NGC 5291N with inclination $55^{\circ}$.  The curves are the same as in Figure 1.}
\label{Fig. 4}
\end{figure}

\vfill\eject

{}

\end{document}